# Linear-response reflection coefficient of the recorder air-jet amplifier

John C. Price

*Department of Physics, 390 UCB, Univ. of Colorado, Boulder, CO 80309*

john.price@colorado.edu

William A. Johnston[a]

*Department of Physics, Colorado State Univ., Fort Collins, CO 80523*

Daniel D. McKinnon[b]

*Department of Chemical Engineering, Univ. of Colorado, Boulder, CO 80309*

(Revised version, Oct. 10, 2015)

[a]Current address: Baker Hughes, Inc., 2851 Commerce Street, Blacksburg, VA 24060.

[b]Current address: Exponent, Inc., 3350 Peachtree Road NE Suite 1125, Atlanta, GA 30326.

**Abstract**

In a duct-flute such as the recorder, steady-state oscillations are controlled by two parameters, the blowing pressure and the frequency of the acoustic resonator. As in most feedback oscillators, the oscillation amplitude is determined by gain-saturation of the amplifier, and thus it cannot be controlled independently of blowing pressure and frequency unless the feedback loop is opened. We open the loop by replacing the recorder body with a waveguide reflectometer: a section of transmission line with microphones, a signal source, and an absorbing termination. When the mean flow from the air-jet into the transmission line is not blocked, the air-jet amplifier is unstable to edge-tone oscillations through a feedback path that does not involve the acoustic resonator. When it is blocked, the air-jet is deflected somewhat outward and the system becomes stable. We are then able to measure the reflection coefficient of the air-jet amplifier versus blowing pressure and acoustic frequency under linear response conditions, avoiding the complication of gain-saturation. The results provide a revealing test of flute drive models under the simplest conditions and with few unknown parameters. The strengths and weaknesses of flute drive models are discussed.

## I. Introduction

In flute-family musical instruments, oscillations of an acoustic resonator are driven by an air-jet, which is formed either by a fixed duct or by the variable geometry of a player's lips. Quantitative models of flute drive mechanisms first appeared in the 1960's and have been continuously developed to the present day.[1,2,3,4,5] The instrument is viewed as a feedback oscillator in which the frequency-determining acoustic resonator, usually a pipe but in some cases a Helmholtz resonator, behaves linearly, or nearly so. Gain is provided by growth of an unstable sinuous mode on the air-jet, which interacts with a sharp edge or lip to create a hydrodynamic flow with a strong periodic component. This flow drives an acoustic oscillation in the resonator, and the resulting acoustic flow disturbs the jet near where it emerges from the duct, thus seeding the unstable sinuous mode and closing the feedback loop. A source of non-linearity is required for any feedback oscillator to maintain a stable amplitude. Fabre *et al*. have argued that vortex shedding at the lip provides this non-linearity in recorders and similar instruments.[6]

Models of flute drive owe much to our understanding of the edge-tone, an oscillation that occurs when an unstable jet strikes a sharp lip in the absence of any acoustic resonance.[7] In the edge-tone phenomenon, the hydrodynamic flow created when the jet interacts with the lip acts directly on the jet at the duct exit, forming a feedback path that supports oscillations. Some recent models of flute drive include both this feedback path, which is called direct hydrodynamic feedback, and feedback through the resonant acoustic flow.[4] The time delay around both feedback loops is due to convection of the unstable mode on the jet at velocities of order 10 m/s.

Flute drive models are often referred to as lumped even though they are not lumped in the strict sense; they involve propagation of sinuous waves on the jet and, in most cases, of acoustic waves in the resonator. However, they are assembled from parts (resonator, jet, various inertances), and all dimensions of the jet are small compared to the acoustic wavelength, a consequence of the convective velocity being much smaller than the speed of sound.

Instruments with a fixed duct geometry, where the blowing pressure is the only relevant jet control parameter, are simpler than those, such as the transverse flute, where the duct is formed by the player's lips. Among the duct-flutes, the recorder has been a popular choice for flute research because the jet is laminar under most playing conditions,[8] a simpler case to model than a turbulent jet. It is also readily available in a range of sizes with highly standardized geometry.

The main goal of the experiments reported here is to test the predictions of a lumped flute drive model under the simplest possible conditions, and with as few free parameters as possible. Steady-state oscillations of the complete instrument necessarily involve non-linearities and gain saturation, effects that are generally beyond the scope of flute drive models, and it is impossible to control the acoustic amplitude and frequency independent of the blowing pressure when the instrument is oscillating. To remove the complicating influence of nonlinearity, we open the acoustic feedback path by replacing the body of the instrument with a waveguide reflectometer, and then measure the linear response of the recorder air-jet amplifier to small amplitude incident waves as a function of frequency and blowing pressure.

The apparatus and methods we have developed for this purpose are described in Section II. In Section III we detail a linearized lumped flute drive model that represents the current state-of-the-art, and that can be compared directly to our data without significant free parameters. In Section IV we present our main results for the reflection coefficient, together with predictions from the flute drive model. We also present results on the playing behavior of the assembled instrument which show how the linear response behavior of the air-jet amplifier is related to the gain-saturated behavior under playing conditions. Implications of our results for flute drive models and possible future experiments are discussed in Section V.

As shown schematically in Fig. 1a, we open the acoustic feedback path by disassembling the instrument at the joint between head and pipe. We measure the complex reflection coefficient of the head $S_h$, defined as the ratio of the complex amplitudes of out-going to in-going pressure waves:

$$S_h(x) = \frac{\varphi_{\text{out}} e^{ikx}}{\varphi_{\text{in}} e^{-ikx}}. \tag{1}$$

A similar expression may be written for the pipe reflection coefficient $S_p$. In both cases we orient the x-axis so that a positive displacement is into the component. Relevant frequencies are assumed low enough so that only the lowest acoustic mode propagates in the cylindrical waveguide, and all periodic quantities are assumed to have $\exp(+i\omega t)$ time dependence, so that $\exp(-ikx)$ with $k$ positive corresponds to a wave moving in the $+x$ direction. The reflection coefficient can be translated to any reference plane location *x* if it is measured at one location and the propagation constant *k* is known.

As indicated in Fig. 1b, the head and pipe can alternately be characterized by the head impedance $Z_h = p/q$ and the pipe impedance $Z_p = p/q$, where $p$ is the acoustic pressure amplitude and $q$ is the acoustic flow (volume/time) amplitude. If an impedance $Z$ and reflection coefficient $S$ are defined at the same reference plane, the relation between them is

$$S = \frac{Z - Z_0}{Z + Z_0}, \qquad Z_0 = \frac{\rho c}{\pi R^2}, \tag{2}$$

where $Z_0$ is the characteristic impedance of the waveguide, $\rho$ and $c$ are the air density and speed of sound, and $R$ is the bore radius at the reference plane. The measurement waveguide has the same bore radius as the head (which is not tapered), so that a single value of $R$ applies throughout the system. The body has an inverse taper, decreasing from the head joint to the foot of the instrument, but this does not play a role in our measurements.

When the system is reassembled so that the in-going wave for the head becomes the out-going wave for the pipe and vice versa, the condition for steady-state motion is $S_h S_p = 1$. For a linear system at real frequencies this can only be satisfied if oscillations are marginally stable. Taking the magnitude squared of this expression and noting that the pipe is passive and satisfies $|S_p|^2 \leq 1$, we conclude that the system is potentially unstable at real frequencies where $|S_h|^2 \geq 1$. This is the condition for pipe-tone oscillations. In terms of the head impedance, the equivalent condition is $\mathcal{R}(Z_h) \leq 0$. (We use $\mathcal{R}(\ )$ and $\mathfrak{I}(\ )$ to represent the real and imaginary parts of expressions.)

When the head reflection coefficient is measured it is connected to a matched transmission line and an impedance-matched signal source, as shown in Fig. 1c. Under these conditions, the head is connected to a system with $S = 0$, and thus, the threshold for instability occurs when $|S_h|^2 = \infty$. This is the edge-tone oscillation threshold, since under these conditions out-going radiation is absorbed and no acoustic resonance is involved. If the system shows edge-tone oscillations when connected to the reflectometer, linear response measurements are not possible. We show below that the edge-tone oscillations depend strongly on whether we allow a portion of the mean flow from the jet to pass into the reflectometer or not.

Much has been learned through experiments on steady-state oscillations of assembled recorders and similar instruments[8,9,10,11,12] where the amplifier is operating in gain-saturated conditions, and the wave amplitude in the bore cannot be controlled independently of other parameters. However, the previous experiments most similar in spirit to the present work were those of Thwaites and Fletcher who studied a two-meter long organ flue-pipe with a turbulent jet.[13] The steady-state standing wave ratio in the pipe was measured using a single microphone that could be translated along the pipe axis, and a sound source and absorbing termination were provided at the passive end of the pipe. The active end was modeled as two admittances in parallel, one of which was fit to the data with the jet turned off. The results were discussed in the context of a controversy about the relative importance of momentum drive versus volume drive mechanisms. Most later authors have cited Coltman's experiments[14] as definitive evidence that momentum drive is small, and have used only the volume drive mechanism, as we do in Section III below. Verge *et al.*[3] tried to probe the linear regime by comparing a time-domain model to the initial transient oscillations that appear when the blowing pressure is abruptly applied to the instrument. They used a recorder-like head geometry and a small pipe of fixed length.

## II. Apparatus and methods

Our apparatus is shown schematically in Fig. 2. The blowing system applies a known steady air pressure to the recorder mouthpiece and measures the air volume flow rate. Compressed air at a gauge pressure of 2-8 bar is regulated to an absolute pressure of 1.11 bar by a precision multi-stage regulator (R1), and the flow rate is measured (F1) at this fixed pressure. A metering valve (V1) then leaks air into a 2 liter settling volume whose pressure is monitored by two digital manometers (M2, M3) covering the gauge pressure range 0-750 Pa. Flexible latex tubing is used to connect the settling volume to the recorder mouthpiece. The volume flow rate of the jet is determined by multiplying the measured flow rate by the ratio of the absolute pressure measured by M1 to the ambient pressure measured by a barometer. Our laboratory in Boulder, Colorado, is at an altitude of 1.6 km where the typical ambient pressure is 840 mbar.

The head used for the measurements reported here is from a Yamaha YRT-304B II tenor recorder, shown in cross-section in Fig. 3. It has a cylindrical bore of radius $R = 12.36$ mm. We chose a tenor head rather than a smaller soprano or alto head because its larger bore reduces the relative importance of microphone perturbations on the reflection coefficient measurements. The planar air-jet is formed by a duct or windway of length $L = 7.20$ cm and width $H = 1.47$ cm (perpendicular to the plane of the figure). Its height is 1.95 mm at the inlet and $h = 1.09$ mm at the exit (see figure inset). The duct is curved about an axis parallel to and below the cylindrical axis of the bore with a radius of curvature of 2.7 cm. At the duct exit are two 45° chamfers that extend a distance $d = 0.7$ mm into the duct. The lip is located a distance $W = 4.83$ mm from the duct outlet, and is aligned with the center of the duct, so that an undisturbed jet would be split approximately equally above and below the lip. The lip has an included angle of 14.8° with the lower surface parallel to the cylindrical axis of the bore, and is curved about an axis parallel to the cylindrical axis following the curvature of the duct.

Returning to Fig. 2, the head is connected to the 40 cm long microphone section with a brass sleeve that maintains an uninterrupted bore. The microphone section, made from CPVC polymer, is bored to match the inside diameter of the recorder head. The amplitude and phase of right- and left-going waves in the waveguide are determined using signals from six electret microphones that are irregularly spaced along the microphone section. (Irregular spacing provides better frequency coverage near the waveguide cut-off frequency, but is not important for the results presented here.) The 6 mm diameter electret elements (Panasonic WM-61a) are cemented into thin-wall aluminum tubes and mounted in O-ring sealed fixtures so that the element inlet ports are flush with the inner bore. The microphone section is connected to a 10 cm long attenuator section, which is filled with three cosmetic cotton balls, providing an absorbing termination and approximately 10 to 20 dB attenuation from 200 Hz to 1500 Hz. A 120 watt compression driver (Selenium D210TI) drives the system through the attenuator.

When a recorder is played, the portion of the jet flow passing below the lip exits the instrument through open tone holes or at the far end of the pipe, which is normally open. In the course of our experiments we found that it is important to control the mean flow $Q_p$ down the pipe when making reflection coefficient measurements. To this end, we connected an 85 liter/minute rotary-vane vacuum pump to the attenuator section through the metering (needle) valve V2. Because the driver is somewhat leaky we also added a thin polymer membrane across the seal between the attenuator and the driver. With this arrangement, the mean pipe flow is zero with the metering valve closed, and can be controllably increased to any fraction of the jet flow $Q_j$ by adjusting V2.

The six electret microphone elements are connected through conventional preamplifiers to the microphone inputs of a digitizer (PreSonus Firepod), which samples the microphone signals synchronously at 44.1 kHz using 24-bit converters. The digitizer's headphone output is used to provide signals for the compression driver. Data is acquired and simultaneous audio output is generated using the ASIO audio drivers and the pa_wavplay audio interface to MATLAB.[15]

A calibration cell is used to correct for relative phase and gain errors of the six microphone channels. The compression driver is removed from the apparatus and connected to a short section of waveguide similar to the attenuator section (Fig. 2), but with a closed end and no absorbing material. The six microphones are mounted in O-ring sealed fixtures close to the closed end, uniformly spaced around the circumference of the waveguide at equal distances from the driver. With this set-up, the same pressure signal can be applied to each microphone channel. The relative phase and amplitude response is recorded for each microphone channel at the same set of frequencies as are used for reflection coefficient measurements, and at comparable acoustic amplitudes. Typically, calibration is performed at the beginning of a data run and then repeated several hours later. Amplitude ratios between microphones are stable to about 2 parts in a 1000, and phase differences are stable to about one milliradian. Better performance can be obtained if calibrations are done more often and care is taken to avoid temperature drift. The main source of relative calibration loss appears to be temperature changes of the electret cells.

The average amplitude temperature coefficient of the six microphones is -0.34% per °C, but this varies across the six channels by ±0.14% per °C. If the temperature of all microphones is changed together, the relative phases vary by ±0.3 milliradians per °C across the six channels. A Brüel & Kjaer type 4231 calibrator is used for absolute pressure calibration of one channel at 1 kHz. Amplitude flatness from 200 Hz to 1500 Hz is believed to be better than ±1 dB, although this plays no role in reflection coefficient measurements.

The reflection coefficient is measured at 100 discrete frequencies, logarithmically distributed from 200 Hz to 1500 Hz. At each frequency, a one second duration sine wave burst is generated by the compression driver. The drive signal amplitude rises and falls smoothly over the first and last 50 ms of the pulse to reduce spectral leakage. Simultaneously, each microphone channel is recorded and the waveforms are windowed with a Hann function, multiplied by sine and cosine functions at the known signal frequency, and then integrated to yield the real and imaginary parts of the signal amplitude. Once the complex signal amplitudes have been determined at each frequency, they are corrected using the calibration data so that the relative magnitudes and phases are those that would be recorded by microphones without relative phase and amplitude errors.

The microphone amplitudes are used together with the known microphone locations to find the wave amplitudes $\varphi_{in}$ and $\varphi_{out}$ and the propagation constant $k$ (see Eqn. 1). The wave amplitudes are fit independently at each frequency but the propagation constant is fit using the entire data set and is constrained to the form:

$$k = \frac{\omega}{c_g} - i\alpha, \qquad c_g = c\left(1 - c\frac{\delta}{\sqrt{\omega}}\right), \qquad \alpha = \beta\sqrt{\omega}, \tag{3}$$

where $c_g$ is the guided wave phase velocity and $\alpha$ is the attenuation constant. The free-space sound velocity $c$ and the parameters $\delta$ and $\beta$ are fit. The form of Eqn. 3 for the attenuation and dispersion is from the Helmholtz-Kirchhoff theory,[16] which predicts $\beta = \delta = 9.5 \cdot 10^{-4} \text{m}^{-1}\text{s}^{1/2}$ for air at standard conditions in a 25 mm I.D. tube. If measurements are made at N frequencies, the six microphone amplitudes correspond to 12N real numbers, while the two wave amplitudes plus the three propagation parameters are 4N+3 real numbers. Thus, the fit is highly over-constrained and the quality of the fit can be examined for each frequency and microphone to test that the apparatus is functioning as expected. With the wave amplitudes and propagation constant known, Eqn. 1 can be used to find the reflection coefficient at any reference plane location $x$.

Performance of the system is checked by replacing the recorder head with a section of waveguide with a closed end. The reflection coefficient with the reference plane at the closed end should then be given by Eqn. 2 with $Z \to \infty$, so that $S = +1$. Typical performance is $S = +1 \pm 0.02 \pm 0.02i$ over the frequency range 200 Hz to 1500 Hz.

**III. Linear model of flute drive**

In this section we describe a linear flute drive model based on the work of Verge *et al.*.[3] Inspired by the ideas of Powell[7] and Coltman,[17] Verge *et al.* represent the interaction of the jet with the lip by two complementary oscillating flow sources $q_j$ and $-q_j$, located above and below the lip (Fig. 4a). The sources drive a potential flow through the mouth of the instrument (the region between the duct exit and the lip), which combines with the acoustic pipe flow $q$ to disturb the jet. Jet disturbance due to $q_j$ corresponds to the direct hydrodynamic feedback mechanism of edge-tone models, while jet disturbance from $q$ represents feedback from the acoustic resonance. A dimensionless linear jet transfer function $J(\omega)$ relates $q_j$ back to the disturbing flows:

$$q_j = (\alpha_p\, q + \alpha_j\, q_j) J(\omega). \tag{4}$$

The constants $\alpha_p$ and $\alpha_j$ account for non-uniformity of the flows across the mouth and would both be equal to one if these flows were taken to be uniform. Verge *et al.* argue that the jet is disturbed mainly by the cross-flow velocity at the duct exit. Using an approximation to the geometry and two-dimensional potential flow calculation, they find $\alpha_p = 2/\pi$ and $\alpha_j = 0.38$.

Figure 4b shows our one-dimensional lumped model of the head impedance. The port at the right labeled by the pipe flow $q$ and pressure $p$ corresponds to the port of the head impedance $Z_h$ shown in Fig. 1b. The drive inertance $M_d$ represents the inertia of the flow driven by the source $q_j$, which is shown (using electronic schematic language) as a current source. The two-dimensional flow calculation mentioned above implies $M_d = 0.88\rho/H$. The inertances $M_1$ and $M_2$ represent the inertia of air just outside and just inside the mouth, while $r$ represents radiation resistance and other acoustic losses in the head. (We find that $r$ is not significant for the data reported here and may be set to zero.) Analysis of this circuit yields the active linear head impedance

$$Z_h = \frac{p}{q} = r + i\omega M_1 + i\omega M_d \frac{1 + (\alpha_p - \alpha_j) J(\omega)}{1 - \alpha_j J(\omega)} + i\omega M_2. \tag{5}$$

If the reference plane is located at the port of $Z_h$, the head reflection coefficient will be

$$S_h = \frac{Z_h - Z_0}{Z_h + Z_0}. \tag{6}$$

At zero blowing pressure, $J(\omega)$ is equal to zero and the head impedance becomes $Z_h = r + i\omega(M_1 + M_d + M_2)$. When comparing the model to our experiments, we fit the total lumped inertance $M = M_1 + M_d + M_2$ to reflection coefficient data at zero blowing pressure. Except for the jet transfer function, all other parameters in Eqns. 5 and 6 are then known.

Our jet model is based on analysis and hot-wire anemometry experiments reported by Nolle,[18] who studied acoustic deflection of planar air-jets at Reynolds and Strouhal numbers similar to those under consideration here, using methods sensitive to very small jet displacements corresponding to linear response. He set out to test a jet deflection model that originates with Fletcher[19] and is based on what he called the flue boundary condition. This supposes that a (small) jet disturbance can be understood as the superposition of an unstable sinuous mode that propagates away from the duct and an acoustic flow transverse to the jet flow direction, with the amplitude and phase of the sinuous mode adjusted to yield zero jet displacement at the duct exit. The displacement amplitude $\xi(s)$ of the disturbed jet is given by

$$\xi(s) = a\left(1 - e^{\mu s}e^{-i\omega s/c_p}\right), \tag{7}$$

where $a$ is the *displacement* amplitude of the acoustic cross-flow, $\mu$ is the growth rate of the transverse sinuous mode of the jet and $c_p$ is the phase velocity of the sinuous mode. Both $a$ and $\xi(s)$ are taken to be positive upward in Fig. 4a. The coordinate $s$ begins at the duct exit and increases along the path of the undeflected jet. The second exponential factor corresponds to a time delay $s/c_p$. In Fletcher's original implementation of the flue boundary condition, he used the factor $\tanh(\mu s)$ rather than $\exp(\mu s)$, and this version was also used by Verge *et al.*. However, Nolle found that Eqn. (7) agreed best with his measurements. Jet models of this kind have not yet been derived from fluid mechanics by any controlled approximation and so must be considered semi-empirical. They have been criticized,[3] but at present no better alternative appears to be available for jets with realistic flow profiles.

We next convert Eqn. 7 into the dimensionless jet transfer function $J(\omega)$. The acoustic amplitude $a$ is related to a uniform acoustic flow $q$ though the mouth by

$$a = \frac{q}{i\omega W H}. \tag{8}$$

The complementary flow sources on the lip due to jet deflection are related to the jet central velocity $U_0$ and the jet displacement by

$$q_j = -\xi(\widetilde{W}) U_0 H, \tag{9}$$

where $\widetilde{W} = W + d$ is the jet travel distance including the chamfer depth at the duct exit. This expression assumes that the undeflected jet strikes the lip at its center and that the deflection is small compared to the jet width. It also assumes that once the jet flow passes the lip edge at $s = W + d$ it is converted into the complementary flow sources $\pm q_j$ with no additional phase delay. Combining Eqns. 7–9 yields

$$J(\omega) = -g \frac{U_0}{W} \frac{1}{i\omega} \left(1 - e^{\mu \widetilde{W}} e^{-i\omega \widetilde{W}/c_p}\right). \tag{10}$$

A numerical pre-factor $g$, which we refer as the jet gain, has been included in the expression as it turns out to be necessary to give a reasonable account of the data presented below.

The growth rate $\mu$ and phase velocity $c_p$ both depend on the transverse velocity profile of the jet. Nolle's hot-wire profile measurements support the assumption that an initially parabolic (Poiseuille) profile at the duct exit quickly evolves into a Bickley velocity profile

$$U(y) = U_0 \text{sech}^2(y/b) \qquad (11)$$

in the free jet, where $y$ is a coordinate transverse to the jet plane (upwards in Fig 4a) and $b$ is the Bickley profile width. Nolle found good agreement between his jet deflection measurements and the growth rate $\mu$ and phase velocity $c_p$ obtained by Mattingly and Criminale[20] in a perturbative analysis of an infinitely long planar jet with the Bickley profile. Nolle repeated and extended the calculations of Ref. 20, and provided results in tabular form, which we have used to generate the spline fits shown in Fig. 5. In the figure, we plot the dimensionless quantities $\mu b$ and $c_p/U_0$ as a function of the Strouhal number $Stb \equiv \omega b/U_0$.

To use the jet model (Eqn. 10) we must determine the Bickley width $b$ and central jet velocity $U_0$. Verge *et al.* argued that the central velocity should remain unchanged from its value at the duct exit, and that the momentum flux of the jet should be conserved in the transition from the Poiseuille to the Bickley profile, which implies $b = 2h/5$. We use this relation. They estimated the central velocity using the Bernoulli formula: $U_0 = \sqrt{2P/\rho}$, where $P$ is the blowing pressure (*i.e.*, the gauge pressure measured by M1 and M2 in Fig. 2). In our apparatus, the jet volume flow $Q_j$ is directly measured (by F1), so we can instead use the relation between the flow and the central velocity for a Poiseuille profile, $Q_j = \frac{2}{3}hU_0$, to find $U_0$. In Fig. 6, we plot the central velocity $U_0$ determined both ways. The two methods converge at high blowing pressures (higher Reynolds numbers) but at low pressures the Bernoulli method is a poor approximation. We therefore use the measured jet flow to determine $U_0$.

The form of $J(\omega)$ in Eqn. 10 and the dimensionless forms of $\mu$ and $c_p$ in Fig. 5 together imply that $J(\omega)$ is a function of geometrical parameters and Strouhal number only. In the present model, spreading of the jet is neglected, and there is no Reynolds number dependence. The role of viscosity is only to establish a Poiseuille profile at the duct exit. Rather than use $Stb$ as defined above, which is appropriate for free jets, we use $St = f\widetilde{W}/U_0$, where $f$ is the acoustic frequency equal to $\omega/2\pi$. Thus $St$ is equal to the jet travel-time across the mouth divided by the acoustic period. Figure 7 shows $J(\omega)$ plotted on the complex plane with $St$ as a parameter along the curve. The values of $St$ where $\Im(J)$ changes sign are indicated.

If the loss term $r$ is neglected, and $g$ is small enough so that edge-tone oscillations do not occur, a consequence of Eqns. 5 and 6 is that $|S_h|^2 > 1$ when $\Im(J) > 0$, and $|S_h|^2 < 1$ when $\Im(J) < 0$. Thus the loop in Fig. 7 between $St = 0.0705$ and $St = 0.2871$ is a region where the air-jet amplifier has gain and pipe-tone oscillations may occur. A second gain region occurs between $St = 0.4964$ and $St = 0.7690$. Using the phase velocity $c_p$ of Fig. 5, the first of these regions corresponds to 0.21 to 3/4 wavelengths of the sinuous mode on the jet, and the second to 5/4 to 7/4 wavelengths. (Because of the constant term in the parentheses in Eqn. 10, these points do not occur at simple fractions of a wavelength when $St$ is small.)

**IV. Results**

Figures 8 and 9 show the measured magnitude and phase of the head reflection coefficient at 100 frequencies from 200 Hz to 1500 Hz, and for nine different blowing pressures from 0 to 300 Pa. The data in these figures were collected with the valve V2 in Fig. 2 closed, so the mean pipe flow $Q_p$ was equal to zero. Under these conditions, the system does not show edge-tone oscillations. It does not oscillate at any of the blowing pressures tested and $|S_h|^2$ does not diverge at any frequency.

The data were collected with an incident sound pressure level of about 50 dB. The results reproduce very closely if the incident acoustic amplitude is increased by a factor of three, showing that the measurements are in the linear response regime. The sound pressure levels we have used are very low compared to the acoustic pressure in the bore when a recorder is assembled and played, which may be as high as 130 dB.[12]

The position of the reference plane and the value of the total lumped inertance $M = M_1 + M_d + M_2$ in the model were adjusted to fit the data at zero blowing pressure (panels labeled 0 Pa in the figures). The model is plotted as a continuous line in both figures. As expected, at zero blowing pressure the head is nearly lossless and $|S_h| = 1.0$ within the accuracy of our measurements. If we express the fit inertance $M$ in terms of a length $L$ using $M = \rho L/WH$, we find $L = 8.9$ mm. Using the formula given in the previous section for $M_d$, this value implies $M_d/M = 0.46$. We adjust the jet gain to $g = 0.145$ to fit the peak height of $|S_h|$ at 200 Pa blowing pressure. Without this reduction of the jet gain, the model would predict edge-tone oscillations for all of the finite blowing pressures shown. All other parameters of the model are determined as described in the previous section, and not fit to the data.

In Fig. 10 we show the boundaries of the regions in the frequency versus blowing-pressure space where our measurements show that the air-jet amplifier has gain (that is, where, $|S_h| > 1.0$), and compare them to the corresponding boundaries predicted by the model. In the model, these boundaries occur at $\Im(J) = 0$ and thus at the fixed values of $St$ labeled in Fig. 7. Therefore, we plot the dimensionless frequency or Strouhal number $St = f\widetilde{W}/U_0$, and the model predicts the horizontal lines in Fig 10. The first region where the air-jet amplifier has gain, in the model from $St = 0.0705 \to 0.2871$, agrees well with the measurements (open circles and stars), except that the boundaries in the data occur at somewhat lower $St$. (For example, in Fig. 8 at 250 Pa, the data show $|S_h| > 1.0$ from 223 Hz to 1100 Hz, while the model predicts $|S_h| > 1.0$ from 296 Hz to 1170 Hz.) We will see below that this large region where the air-jet amplifier has gain corresponds well to the normal playing region of the instrument. The second region where $|S_h| > 1.0$ can only be seen in the data at blowing pressures below 100 Pa. In the model it is predicted to occur at $St = 0.4964 \to 0.7690$, but in the data it again occurs at lower frequencies corresponding to $St = 0.43 \to 0.60$ (crosses and diamonds).

In Fig. 11, we show what happens at a blowing pressure of 80 Pa when the valve V2 is adjusted to increase the mean pipe flow. The panels are labeled by $Q_p/Q_j$, the ratio of the mean pipe flow to the jet flow. For $Q_p/Q_j = 0$, the response at 80 Pa looks very similar to the 70 Pa data in Fig. 8. However, when $Q_p/Q_j$ is increased to 0.069, the peak height of $|S_h|$ near 500 Hz increases by a factor of 2.5, corresponding to an increase of the jet gain $g$ from 0.145 to approximately 0.20. When $Q_p/Q_j$ is increased further, the head begins to oscillate at a frequency very close to the position of the peak in $|S_h|$ at zero mean pipe flow. If $Q_p/Q_j$ is increased even further to 1.00, the oscillations cease, and the results are similar to those at $Q_p/Q_j = 0$. We thus observe edge-tone oscillations in the intermediate region where the pipe flow is greater than zero but less than the jet flow. We conclude that the peaks in $|S_h|$ at zero mean pipe flow (Fig. 8) are nascent or sub-threshold edge-tone oscillations.

## V. Discussion

Putting aside for a moment the value of the jet gain *g*, the flute drive model of Section III shows surprising agreement with the reflection coefficient data presented in Figures 8 and 9. As far as we are aware, no similar data has been reported previously, so there is no sense in which flute drive models have been refined to fit data of this kind. Besides *g*, the only quantity fit to the data is the value of the total lumped inertance *M*, and it is fit only to the data at zero blowing pressure. The locations of the gain peaks, and the variation of their heights and widths with blowing pressure, are accurately predicted. The phase variations show in Figure 9 have the correct sign and magnitude, demonstrating that the locations of the poles in the complex frequency plane are correctly accounted for.

Although the model shows good qualitative agreement with most features of our data, the predicted gain boundaries shown in Fig. 10 occur at frequencies that are too high, especially at higher Strouhal numbers, suggesting that either the model needs additional time delays or that the modeled phase velocity is too high. The shapes of the predicted reflection coefficient peaks in the second gain region are qualitatively correct and they have the correct phase behavior, but their amplitudes are too large (for example see Fig. 8, 70

Pa, 1000-1500 Hz), suggesting that the modeled sinuous mode growth rate $\mu$ may be too large at higher frequencies.

The most significant disagreement between our measurements and the lumped model is the small value $g = 0.145$ of the jet gain needed to fit the data. When the mean pipe flow $Q_p$ is not blocked, we expect that the jet flow is split equally above and below the lip, and the jet velocity at the lip edge should be equal to the central velocity $U_0$. This is the condition assumed in deriving the model, corresponding to $g = 1$. However, when the pipe flow is blocked, the jet flow split below the lip must eventually exit the system through the mouth. This creates a mean cross-flow that will deflect the jet outward, so that it strikes the lip where its velocity is less than $U_0$, decreasing $g$. On the other hand, when we adjust the pipe flow so that it equals the jet flow, the portion of the jet flow split to the outside must enter as a cross-flow of opposite sign through the mouth, and we expect the jet to be deflected inward, again decreasing $g$ by a similar factor. This picture is qualitatively consistent with the data in Fig. 11, where we see similar reflection coefficients for $Q_p/Q_j = 0$ and $Q_p/Q_j = 1$, but edge-tone oscillations at intermediate values. Unfortunately, we cannot measure $g$ reliably when the jet is undeflected and edge-tone oscillations are present, because the system is not then behaving linearly.

We can try to include jet deflection in the model by replacing the jet velocity at the lip edge $U_0$ in Eqn. 9 by the Bickley profile (Eqn. 11) and interpreting $y$ in that equation as the mean deflection of the jet at the lip edge. This implies $g = \text{sech}^2(y/b)$. Jet deflection by a mean cross-flow has been discussed by Nolle[18] for organ pipes, which are sometimes stopped at the passive end, and by Fletcher[21] for panpipes, which are always stopped. Nolle's calculation is based on the idea that the transverse momentum flux in the jet at the position of the lip should equal the transverse momentum flux entering the jet from the cross-flow. Using this idea, and setting the flow split under the lip equal to the cross-flow, we find $g = 0.58$ for a Bickley jet with blocked pipe flow. Spreading of the jet due to viscosity will also reduce the velocity at the lip and consequently $g$, but a calculation shows that this reduction should be less than 10 percent.[22]

Since these effects are not large enough, we consider other ways that the jet gain $g$ might be reduced. Nolle's hot-wire experiments used a square duct exit, but the recorder head we are using, in common with most recorders, has chamfers at the duct exit (Fig. 3, inset). Ségoufin et al.,[11] Blanc et al.[23] and de la Cuadra,[24] have shown that chamfers play an important role, and a theory for this effect was proposed.[23] Jet visualization studies[24] showed that neither the phase velocity nor the growth rate of sinuous modes depend strongly on the presence of chamfers, but the amplitude of the jet disturbance for a given acoustic cross-flow is about a factor of two smaller for 45° chamfers than for a square exit, corresponding to a decrease of $g$ by a factor of two. This observation is in rough agreement with the Blanc et al. theory. Experiments on edge tones by Ségoufin et al.[25] showed that the threshold for oscillation moves to lower jet velocities when chamfers are added, again consistent with a decrease of $g$.

The combined effects just discussed do not seem large enough to account for the value of $g$ suggested by our data, although we feel that the effects of chamfers needs to be studied more in experiments. Flute drive models incorporate a highly idealized picture of the interaction of the jet with the lip, in which all of the time-varying flow split by the lip immediately appears in the sources $\pm q_j$. This might have to be improved. Jet curvature seems not to have been addressed in the literature, but it could be important, and might lead to a smaller value of the growth rate $\mu$ than we have assumed, which would have an effect similar to a reduction of $g$.

It is not clear *a priori* that the linear-response measurements reported here should bear any relationship to the behavior of the instrument when it is actually played, since strong nonlinearities appear when the air-jet amplifier saturates. To explore this issue, we display in Fig. 12 the normal playing region of the assembled instrument in frequency versus blowing-pressure parameter space, and also plot on the same axes the gain boundary data from Fig. 10. We see that the high-frequency boundary of the first gain region (stars in both Fig. 12 and Fig. 10) closely bounds the playing region. (The low-frequency boundary of the first gain region is below the region shown in Fig. 12.) This suggests that the region where the amplifier has gain at realistic acoustic amplitudes may not be very different from the region observed in linear response. As shown in Fig. 11, the frequency where the edge-tone instability occurs is only slightly below the high-frequency gain boundary. Thus, at a given value of blowing pressure, the instrument is not played near the edge-tone instability where the air-jet amplifier's linear-response gain is highest, but rather at lower frequencies where the gain is lower but still positive.

Under realistic conditions where the mean pipe flow is not blocked, the recorder air-jet amplifier is unstable to edge-tone oscillations over the whole blowing-pressure range explored. On the other hand, when the instrument is actually played, steady-state oscillations do not occur at the frequency of the edge-tone instability, but rather at a lower frequency determined by the pipe resonance, and edge-tone oscillations are suppressed. This implies that edge-tone oscillations should disappear if reflection measurements are made at higher acoustic amplitudes, beyond linear response, even if mean pipe flow is not blocked. Preliminary experiments do show that the edge-tone oscillation disappears at incident amplitudes 20-30 dB below realistic playing amplitudes. Future experiments should address this issue, and also the gain-saturation behavior of the air-jet amplifier.

## VI. Conclusions

We have developed a waveguide method that can be used to study the active reflection coefficient of the recorder air-jet amplifier with independent control of the incident wave amplitude and frequency. If the mean pipe flow is not blocked, the amplifier is unstable to edge-tone oscillations, making linear response measurements impossible. However, when the pipe flow is blocked, deflection of the air-jet away from the lip reduces gain, edge-tone oscillations are suppressed, and the reflection coefficient can be measured in linear response. A comparison of a linear lumped model to the data shows surprisingly good qualitative agreement, but the jet gain in the model has to be reduced by a large factor to correspond with the data. Some reduction of the jet gain is expected when the pipe flow is blocked, but this does not seem to account for the entire effect, and it is not clear at present if some significant extension of the model may be needed.

The region in frequency versus blowing-pressure space where the air-jet amplifier has gain in linear response corresponds well to the normal playing region of the assembled instrument, suggesting that the same region may apply when realistic acoustic amplitudes are present in the bore of the instrument. The air-jet amplifier is unstable to edge-tone oscillations under normal playing conditions, but these oscillations are suppressed and replaced by pipe-tone oscillations in steady-state when the instrument is played.

Finally, we note that several authors have recently reported computational fluid dynamics (CFD) studies of complete recorders and organ pipes that appear to realistically capture jet formation, jet instability, and the interaction of the jet with the lip and the bore resonance.[26,27,28,29] This development raises the question of what observables could be extracted from CFD experiments for comparison with both theory and experiments. We suggest that modeling the head reflection coefficient by CFD may be simpler than modeling the entire instrument, and doing so might have the same advantages as in experiments, namely, independent control of the acoustic frequency, amplitude and blowing pressure.

**References**

1. B. Fabre and A. Hirschberg, "Physical Modeling of Flue Instruments: A Review of Lumped Models," Acustica **86**, 599-610 (2000).

2. L. Cremer and H. Ising, "The Self-Excited Vibrations of Organ Pipes," Acustica **19**, 142-153 (1967).

3. M.-P. Verge, R. Caussé, B. Fabre, A. Hirschberg, A.P.J Wijnands, A. van Steenbergen, "Jet oscillations and jet drive in recorder-like instruments," Acta Acustica **2**, 403-419 (1994).

4. M.-P. Verge, "Aeroacoustics of confined jets with applications to the physical modeling of recorder-like instruments," Ph.D. thesis, Technische Universiteit Eindhoven (1995).

5. R. Auvray, A. Ernoult, B. Fabre, P.-Y. Lagrée, "Time-domain simulation of flute-like instruments: Comparison of jet-drive and discrete-vortex models," J. Acoust. Soc. Am. **136**, 389-400 (2014).

6. B. Fabre, A. Hirschberg, A.P.J Wijnands, "Vortex shedding in the steady oscillation of a flue organ pipe," Acustica **82**, 863-877 (1996).

7. A. Powell, "On the edgetone," J. Acoust. Soc. Am. **33**, 395-409 (1961).

8. F. Blanc, B. Fabre, N. Montgermont, P. de la Cuadra, A. Almeida, "Scaling of flute-like instruments: an analysis from the point of view of the hydrodynamic instability of the jet," Acta Acustica **96**, 642-653 (2010).

9. J. Martin, "The Acoustics of the Recorder," Moeck Verlag, Celle, Germany (1994).

10. N.H. Fletcher, L.M. Douglas, "Harmonic generation in organ pipes, recorders, and flutes," J. Acoust. Soc. Am. **68**, 767-771 (1980).

11. C. Ségoufin, B. Fabre, M.-P. Verge, A. Hirschberg, A.P.J. Wijnands, "Experimental Study of the Influence of the Mouth Geometry on Sound Production in a Recorder-like Instrument: Windway Length and Chamfers," Acustica **86,** 649-661 (2000).

12. M.-P. Verge, B. Fabre, A. Hirschberg, A.P.J. Wijnands, "Sound production in recorderlike instruments. I. Dimensionless amplitude of the internal acoustic field," J. Acoust. Soc. Am. **101**, 2914-2924 (1997).

13. S. Thwaites and N.H. Fletcher, "Acoustic admittance of organ pipe jets," J. Acoust. Soc. Am. **74**, 400-408 (1983).

14. J.W. Coltman, "Momentum transfer in jet excitation of flute-like instruments," J. Acoust. Soc. Am. **69**, 1164-1168 (1981).

15. URL http://www.mathworks.com/matlabcentral/fileexchange/4017-pa-wavplay (Date last viewed 10/14/15).

16. J.P.M Trusler, "Physical Acoustics and Metrology of Fluids," Taylor and Francis, New York (1991).

17. J. W. Coltman, "Jet drive mechanisms in edge tones and organ pipes," J. Acoust. Soc. Am. **60**, 725-733 (1976).

18. A.W. Nolle, "Sinuous instability of a planar air jet: Propagation parameters and acoustic excitation," J. Acoust. Soc. Am. **103**, 3690-3705 (1998).

19. N.H. Fletcher, "Sound production by organ flue pipes," J. Acoust. Soc. Am. **60**, 926-936 (1976).

20. G.E. Mattingly and W.O. Criminale, "Disturbance Characteristics in a Plane Jet," Phys. Fluids **14**, 2258-2264 (1971).

21. N.H. Fletcher, "Stopped-pipe wind instruments: Acoustics of the panpipes," J. Acoust. Soc. Am. **117**, 370-374 (2005).

22. D.J. Tritton, "Physical Fluid Dynamics," 2$^{nd}$ Ed., Clarenden Press, Oxford (1988).

23. F. Blanc, V. François, B. Fabre, P. de la Cuadra, P.-Y. Lagrée, "Modeling the receptivity of an air jet to transverse acoustic disturbance with application to musical instruments," J. Acoust. Soc. Am. **135**, 3221-3230 (2014).

24. P. de la Cuadra, "The sound of oscillating air jets: Physics, modeling and simulation in flute-like instruments," Ph.D. thesis, Stanford University (2005). https://ccrma.stanford.edu/~pdelac/research/MyPublishedPapers/Thesis.pdf. (Date last viewed 10/14/15).

25. C. Ségoufin, B. Fabre, L. de Lacombe, "Experimental investigation of the flue channel geometry influence on edge-tone oscillations," Acustica **90**, 966-975 (2004).

26. N. Giordano, "Simulation studies of a recorder in three dimensions," J. Acoust. Soc. Am. **135**, 906-916 (2014).

27. M. Miyamoto, Y. Ito, T. Iwasaki, T. Akamura, K. Takahashi, T. Takami, T. Kobayashi, A. Nishida, M. Aoyagi, "Numerical Study on Acoustic Oscillations of 2D and 3D Flue Organ Pipe Like Instruments with Compressible LES," Acta Acustica united with Acustica **99**, 154-171 (2013).

28. H. Yokoyama, A. Miki, H. Onitsuka, A. Iida, "Direct numerical simulation of fluid-acoustic interactions in a recorder with tone holes," J. Acoust. Soc. Am. **138**, 858-873 (2015).

29. J. Fischer, S. Bergweiler, M. Abel, "The Arnold-Tongue of Coupled Acoustic Oscillators," arXiv:1311.5797v1.
29. J. Fischer, S. Bergweiler, M. Abel, "The Arnold-Tongue of Coupled Acoustic Oscillators," arXiv:1311.5797v1.

**Figure Captions**

Figure. 1. a) The recorder is disassembled at the joint between the head and the pipe. b) The head and pipe may be represented by impedances $Z_h$ and $Z_p$. c) For reflection coefficient measurements, the head is connected to a matched transmission line and signal source.

Figure 2. Schematic of apparatus showing blowing system, recorder head, phase-coherent microphone section, attenuator, compression driver, and metering valve connected to vacuum pump. R1: ControlAir 100-CB 2-60 psi precision regulator, M1: Goodman-Kleiner 0-300 mm Bourdon gauge, F1: EKM Metering EKM-PGM.75 gas flow meter, V1: Parker 4Z(A)-NLL-V-SS-K metering valve, M2: Dwyer DM-2004-LCD manometer, M3: Dwyer DM-2006-LCM manometer, V2: Swagelok SS-2MG4 metering valve.

Figure 3. The geometry of the Yamaha YRT-304B II tenor recorder head shown in cross-section.

Figure 4. a) A sinuous mode propagating on the jet is converted to oscillating flow sources $q_j$ and $-q_j$ on the lip. The flow due to these sources and the acoustic pipe flow $q$ act back on the jet near the duct exit. b) Lumped model showing the inertance $M_d$ driven by $q_j$ and inertances $M_1$ and $M_2$ representing air just outside and just inside the mouth. The resistance $r$ represents radiation or other losses.

Figure 5. The growth rate $\mu$ (solid line) and phase velocity $c_p$ (dashed line) of sinuous modes on the jet as a function of the Strouhal number $Stb \equiv \omega b/U_0$, after Refs. 18 and 20.

Figure 6. The central jet velocity inferred from the blowing pressure and the Bernoulli law (diamonds) and from the measured jet flow (open circles) assuming a Poiseuille profile at the duct exit. The solid line is a spline fit to the open circles. This curve is used to determine the central jet velocity when comparing the lumped model to data.

Figure 7. The real and imaginary parts of the jet transfer function $J(\omega)$ on the complex plane for $g = 1$, with the Strouhal number $St = f\widetilde{W}/U_0$ as a parameter along the curve. The imaginary part changes sign at the open circles, which are labeled with values of the Strouhal number.

Figure 8. The head reflection coefficient magnitude at zero mean pipe flow for nine different blowing pressures from 0 Pa to 300 Pa. The solid line is the lumped model discussed in Sec. III. The total lumped inertance $M$ and the jet gain $g$ have been adjusted to fit the data. Only every $5^{th}$ data point is included in the 0 Pa plot so that the solid line representing the model is not hidden.

Figure 9. The head reflection coefficient phase at zero mean pipe flow. Only every $5^{th}$ data point is included in the 0 Pa plot so that the solid line representing the model is not hidden.

Figure 10. Boundaries of the regions in the frequency versus blowing-pressure parameter space where the amplifier has gain. $St = f\widetilde{W}/U_0$ is used as a dimensionless frequency and the solid lines are predictions of the lumped model. Open circles and stars: lower and upper boundaries of first gain region. Crosses and diamonds: lower and upper boundaries of second gain region.

Figure 11. Reflection coefficient magnitude for several values of the pipe flow to jet flow ratio $Q_p/Q_j$. At $Q_p/Q_j = 0.465$, the head oscillates at 512 Hz. The model is plotted with $g = 0.145$ at $Q_p/Q_j = 0$ and with $g = 0.20$ at $Q_p/Q_j = 0.069$.

Figure 12. Frequency versus blowing pressure for the 15 natural notes from C4 to C6 of the assembled instrument at 22 C. Solid circles are plotted at the blowing pressure for in-tune notes assuming an equal-tempered scale with A4=440 Hz. Open circles are 10, 20, 30 and 40 cents above and below the in-tune pitch. Upward and downward pointing arrows indicate that the pitch changed discontinuously within the range ±40 cents. 750 Pa was the highest blowing pressure available. Stars, crosses and diamonds are the same data as plotted in Fig. 10 and the solid lines are a fit to these points.

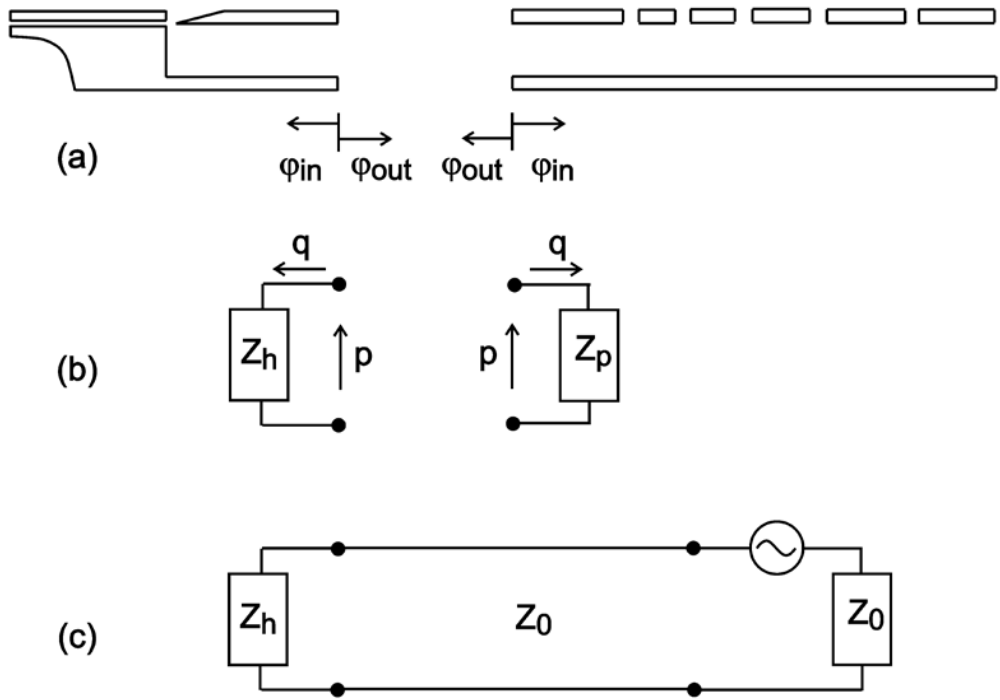

**Figure 1**

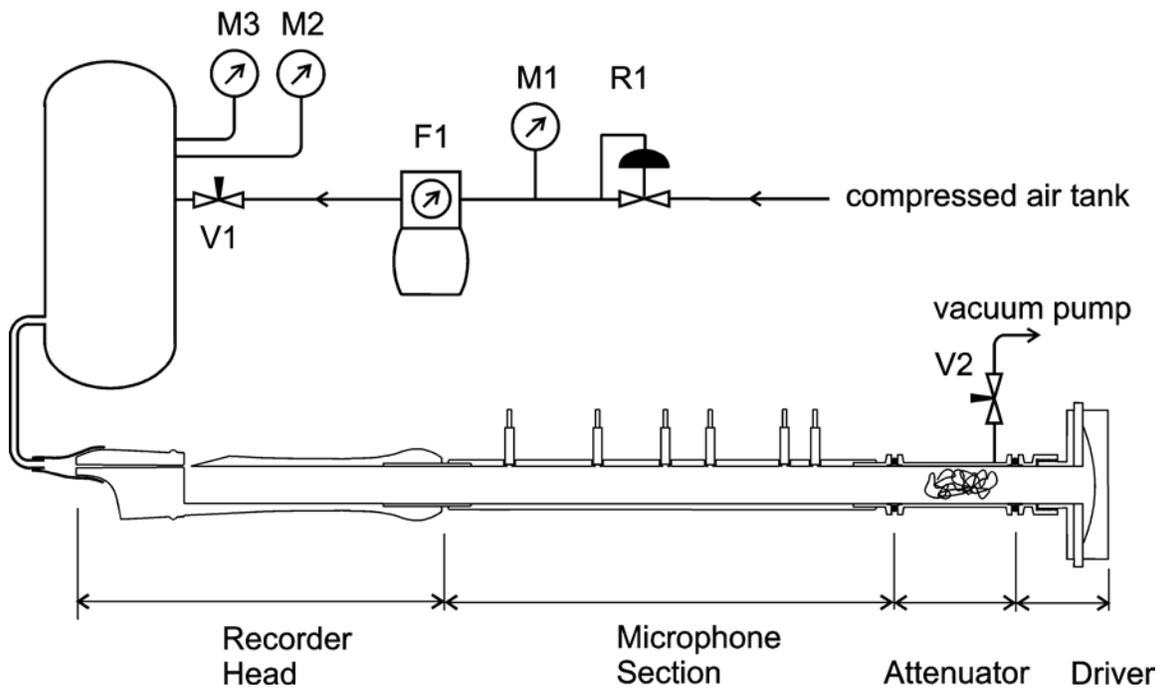

**Figure 2**

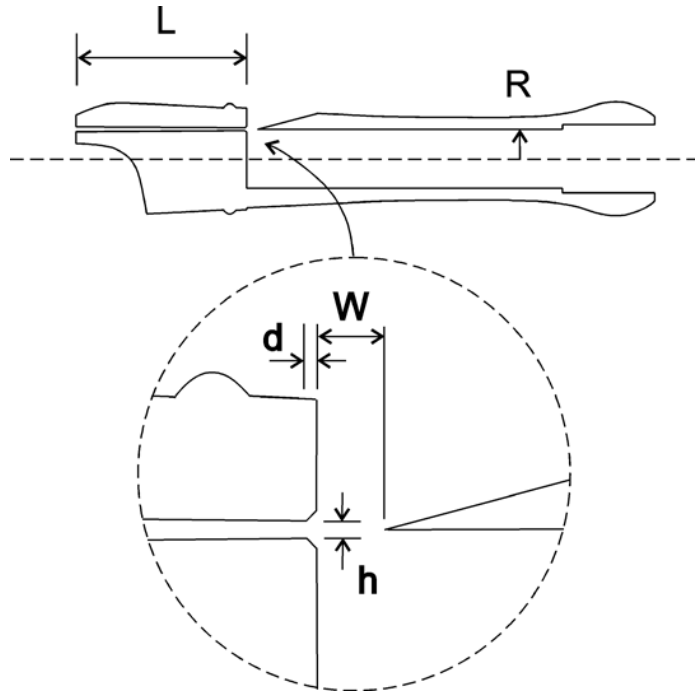

**Figure 3**

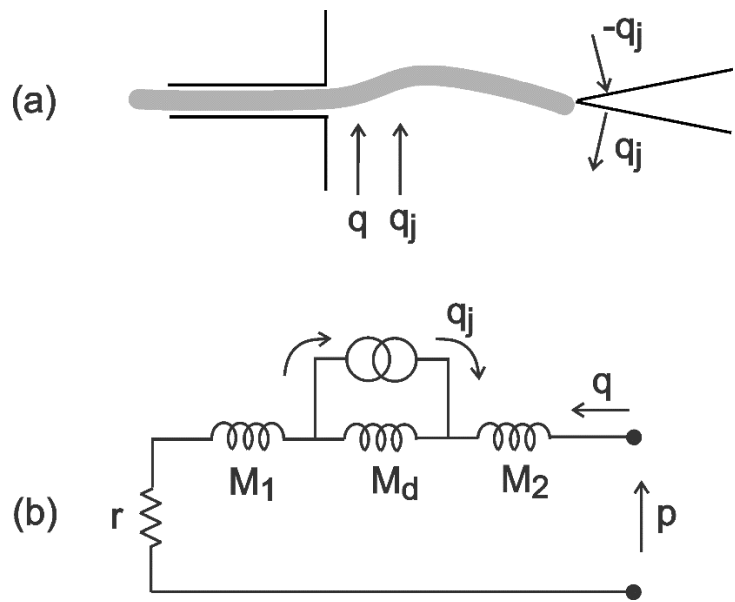

**Figure 4**

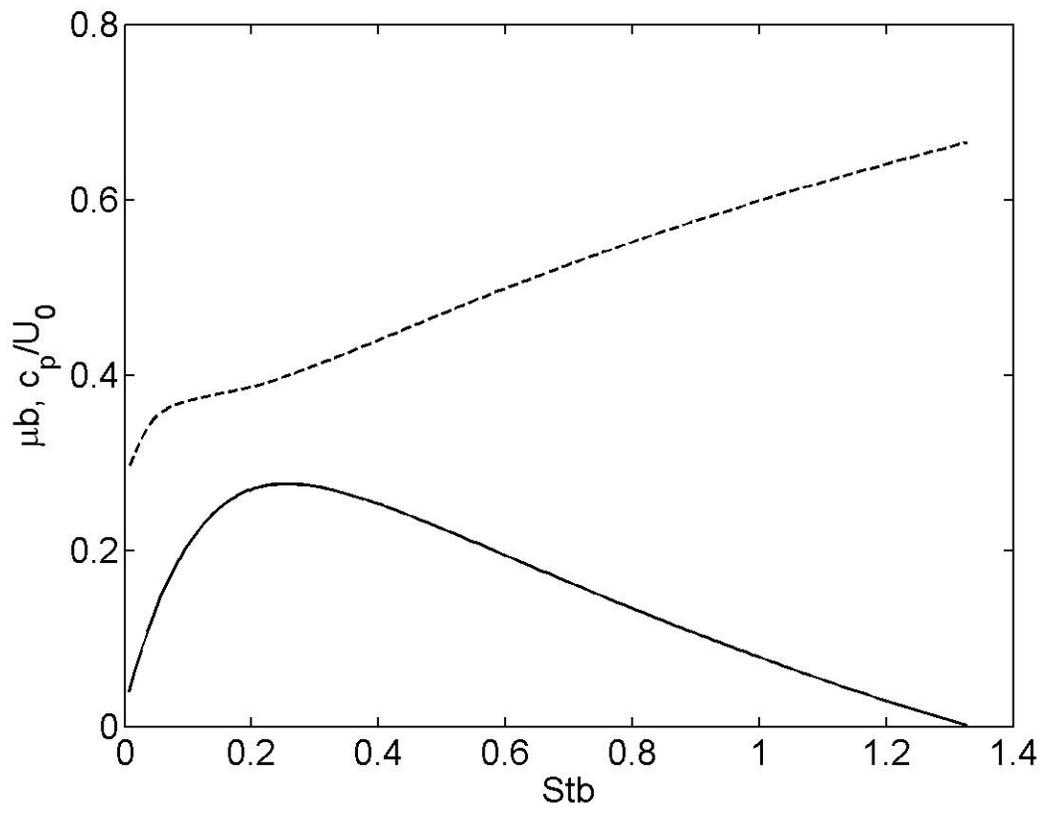

**Figure 5**

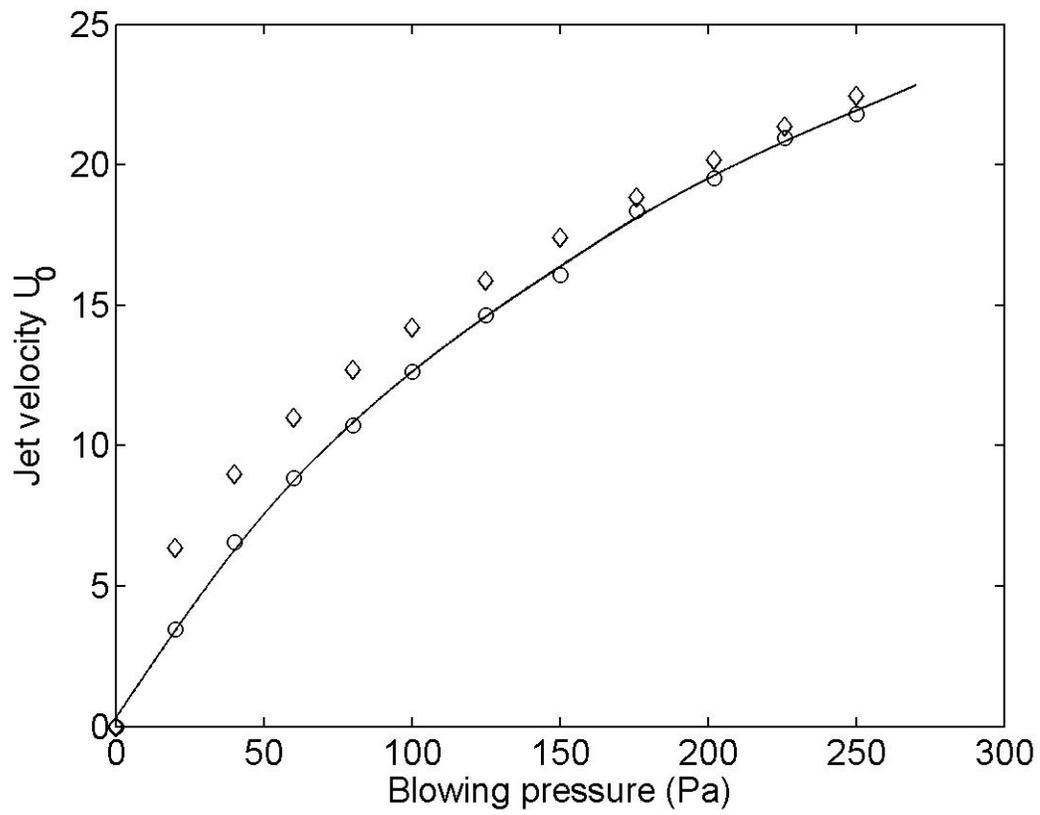

**Figure 6**

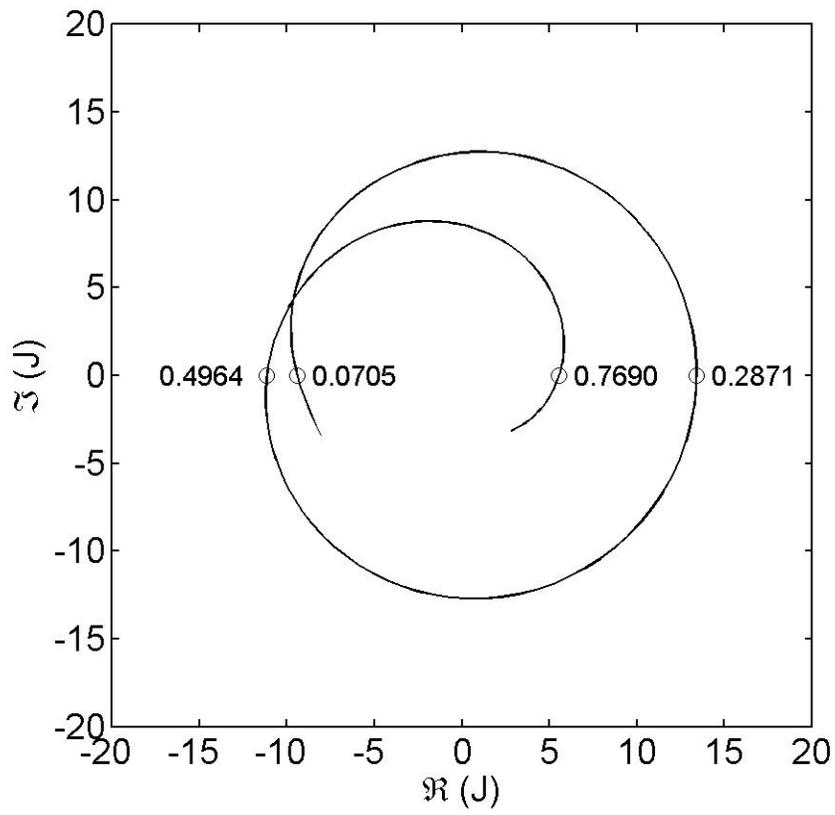

**Figure 7**

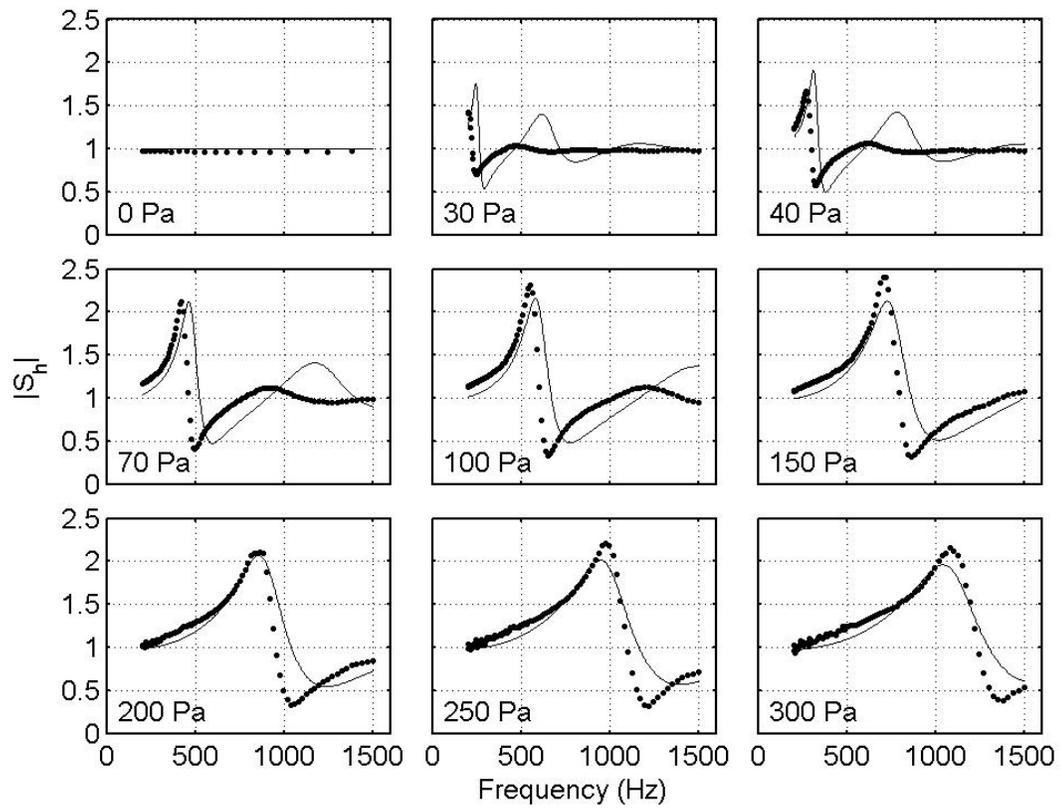

**Figure 8**

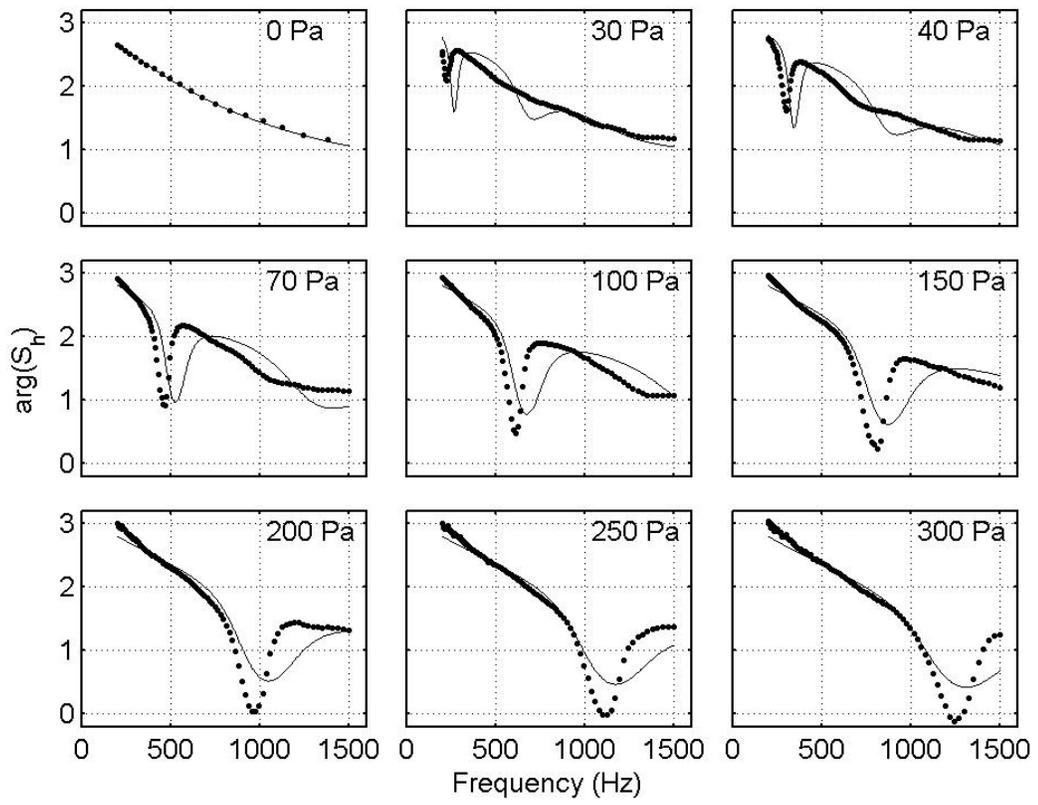

**Figure 9**

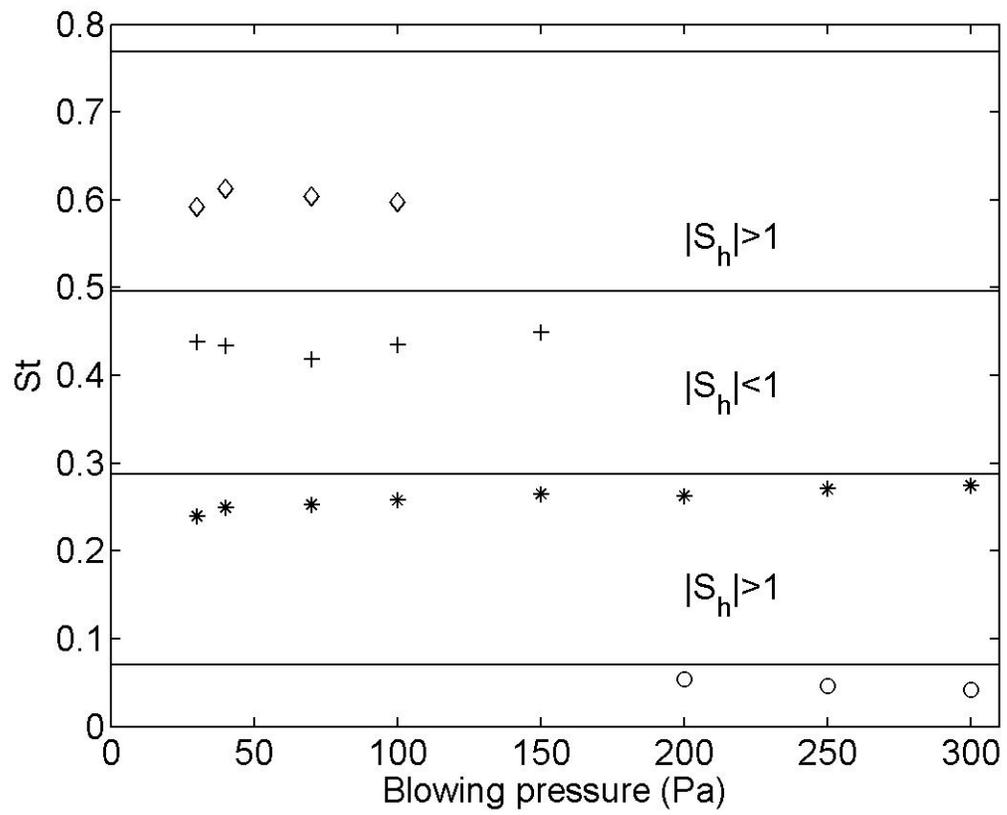

**Figure 10**

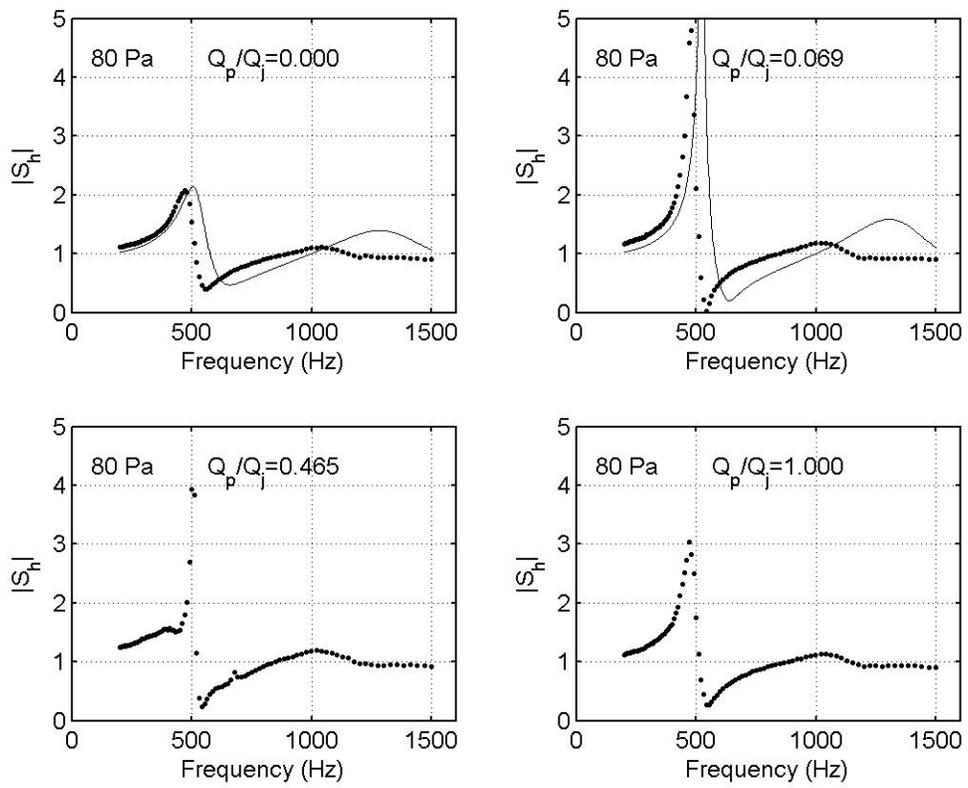

**Figure 11**

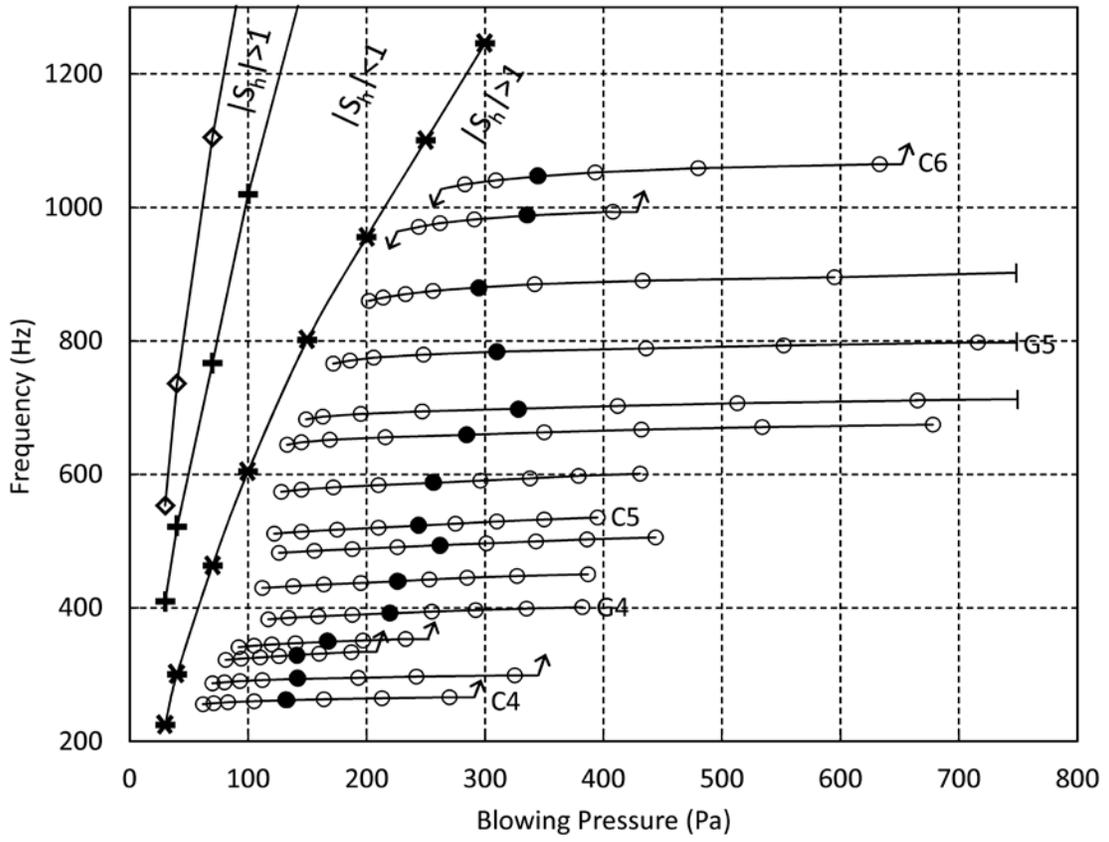

**Figure 12**